\documentclass[10pt,twocolumn]{article}  
\usepackage{amsmath}
\usepackage{url}
\usepackage{graphicx}
\usepackage[small,normal,bf,up]{caption2}
\usepackage{verbatim}

\begin{document}

\title{Nestedness in mutualistic networks\footnote{This manuscript was originally submitted as a Brief Communication to Nature arising from James {\em et al.}, {\em Nature} {\bf 487,} 227-230 (2012)}}

\author{\small Rudolf P. Rohr$^1$, Miguel A. Fortuna$^1$, Bartolo Luque$^2$ \& Jordi Bascompte$^1$\footnote{To whom correspondence should be addressed. E-mail: rudolf.rohr@ebd.csic.es, bascompte@ebd.csic.es}\\
\small $^1$Integrative Ecology Group Estaci\'on Biol\'ogica de Do\~nana (EBD-CSIC) \\ \small c/ Am\'erico Vespucio s/n, Sevilla 41092, Spain \\
\small $^2$Departamento de Matem\'atica Aplicada y Estad\'{\i}stica E.T.S.I. Aeron\'auticos \\ \small Universidad Polit\'ecnica de Madrid Plaza Cardenal Cisneros 3, \\ \small Madrid 28040, Spain}

\date{January 16, 2013}

\vspace{-4 in}

\maketitle

\vspace{0.1 in}

James {\em et al.}$^1$ presented simulations that apparently falsify the analytical result by Bastolla {\em et al.}$^2$, who showed that nested mutualistic interactions decrease interspecific competition and increase biodiversity in model ecosystems. This contradiction, however, mainly stems from the incorrect application of formulas derived for fully connected networks to empirical, sparse networks.

Bastolla {\em et al.}$^2$ showed analytically that a model of mutualistic networks has two solutions, the weak regime and the strong regime.  The former leads to a stable community up to a given threshold in the strength of mutualistic interactions. Beyond this threshold, the weak solution becomes unstable and the strong regime becomes stable. 

James {\em et al.}$^1$ used what they thought to be a set of parameters within the weak regime.  However, this set of parameters was determined by using the formula for a fully connected network. For empirical networks, the condition of the weak regime is that the effective competition matrix is positive definite, which is the case in only 22 out of the 59 studied networks. Thus, the simulations of James {\em et al.}$^1$ occur mainly in another regime than the one assumed by the theoretical development of Bastolla {\em et al.}$^2$ For example, the notion of effective competition is not even defined in the strong regime, already casting doubts on the appropriateness of their comparison. 

The same confusion of formulas that stand only for fully connected networks lead James {\em et al.}$^1$ not to include the effect of the productivity vector variability on biodiversity. Consequently, these authors miscalculated the upper bound of biodiversity and missed the effects of nestedness 

Indeed, the general formula for the maximum biodiversity (here extended for asymmetric effective competition matrices to mimic James {\em et al.}'s parameter sampling procedure; see Methods) is not only a function of the effective competition, but also of the variability of the effective productivity vector.  This effective productivities are similar across species in fully connected networks, since they have the same number of interactions. Real networks, however, show a strong heterogeneity in the number of interactions across species, and therefore in the components of their productivity vectors.  Using the appropriate mathematical formulas (see Methods), we confirm that the predicted upper bound of biodiversity is much closer to the observed values. Specifically, the predicted value is 2 to 10 times larger than the observed value as opposed to 1000 times larger as computed by James {\em et al.} (see Figure 1a). 

\begin{figure}[hbtp]
\centerline{\includegraphics[width=0.9\linewidth]{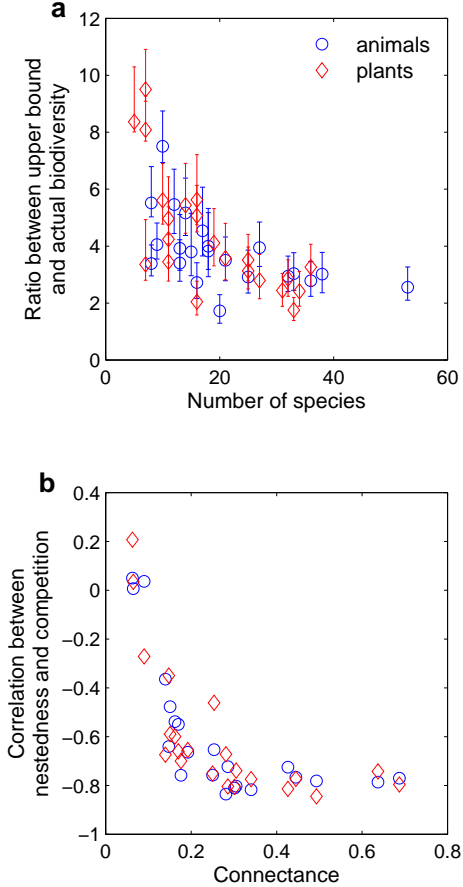}}
\caption{Ratio between the upper bound and actual biodiversity (a) and correlation between the effective competition and nestedness (b) using the generalized analytical expressions for structured networks. Parameter values as in Ref. 1 applied to the 22 mutualistic networks that are in the weak regime.  Growth rates are sampled uniformly in $[0.8,1.4]$ to guarantee that the variability in the effective productivity vector is not too strongly correlated to nestedness. 1000 replicates are used. }
\end{figure}

Regarding the effect of nestedness on biodiversity, our calculation with the extended correct formula (see Methods) shows that the effective competition---which is inversely related to biodiversity--- is almost always negatively correlated to nestedness (Figure 1b).  Similarly, using a multilinear analysis (see Methods) we explicitly disentangle the effects of nestedness and those of the effective productivity vector.  We have done this independently for each network with the goal of removing the confounding effects of network size and connectance. This results in nestedness being always significantly and positively related to biodiversity. 

In summary, the apparently contradictory results by James {\em et al.}$^1$ can be turned upside-down if one applies the correct formulas for structured networks. Their approach would be like trying to prove that Pythagoras' theorem is false for the counter example of a non-right triangle. Both the expanded analytical results and the correct analysis of the simulations here shown confirm that the observed architecture of mutualistic networks indeed increases biodiversity.

\vspace{0.2 in}

{\bf Methods}

We generalize the effective competition formula as $\displaystyle \rho  = \frac{1 - \hat{\sigma}}{ \hat{\sigma} (S-1) + 1}$, where $\displaystyle \hat{\sigma} = \frac{\frac{1}{S-1} \sum_{k=2}^S \sigma^k}{\sigma^1}$ and $\sigma^i$ are the singular values of the normalized effective competition matrix. The upper bound of biodiversity is generalized as $\displaystyle  \frac{1-\rho}{\rho} \left(\frac{\tilde{\sigma}}{\Delta} - 1 \right)$, where $ \displaystyle \tilde{\sigma} = \frac{\sigma^2}{\frac{1}{S-1} \sum_{k=2}^S \sigma^k}$ and $\displaystyle \Delta = \frac{1 - \cos(\theta_1) \cos(\theta_2)}{\cos(\theta_1) \cos(\theta_2)}$. $\theta_1$ ($\theta_2$) is the angle between the effective productivity vectors and the first left-(right)-singular vector of the effective competition matrix. 
We relate biodiversity with nestedness and productivity as: $\displaystyle \log(\text{biodiversity}) \sim \log(\eta) + \log(1/\Delta)$.

\vspace{0.2 in}

{\large \bf References}

{\small

\begin{description}

\item 1. James, A., Pitchford, J.W. \& Plank, M.J. Disentangling nestedness from models of ecological complexity, {\em Nature} {\bf 487,} 227-230 (2012).

\item 2. Bastolla, U., Fortuna, M.A., Pascual-Garc\'ia, A., Ferrera, A., Luque, B. \& Bascompte, J. The architecture of mutualistic networks minimizes competition and increases biodiversity, {\em Nature} {\bf 458,} 1018-1020 (2009).

\end{description}
}

\end{document}